# Epitaxial thin films of the multiferroic double perovskite $Bi_2FeCrO_6$ grown on (100)-oriented $SrTiO_3$ substrates: Growth, characterization, and optimization

Riad Nechache,[1] Catalin Harnagea,[1] Louis-Philippe Carignan,[2] Olivier Gautreau,[1] Lucian Pintilie,[4,b] Mangala P. Singh,[3] David Ménard,[2] Patrick Fournier,[3] Marin Alexe,[4] and Alain Pignolet[1]

[1] Énergie, Matériaux et Télécommunications, INRS, 1650, boulevard Lionel-Boulet, Varennes, Québec J3X 1S2, Canada
[2] Département de Génie Physique, École Polytechnique de Montréal, P.O. Box 6079, Station Centre-ville,
Montréal, Québec H3C 6A7, Canada
[3] Département de Physique, Université de Sherbrooke, Sherbrooke, Québec J1K 2R1, Canada
[4] Max Planck Institute of Microstructure Physics, Weinberg 2, D-06120 Halle/Saale Germany
[b] Present address: National Institute of Materials Physics, Atomistilor 105bis, Magurele, Ilfov 077125, Romania. Electronic mail:pintilie@infim.ro.

## ABSTRACT

The influence of the deposition pressure $PO_2$ and substrate temperature $T_S$ during the growth of $Bi_2FeCrO_6$ thin films grown by pulsed laser deposition has been investigated. It is found that the high volatility of Bi makes the deposition very difficult and that the growth of pure Bi2FeCrO6 thin films on $SrTiO_3$ substrates is possible only in a narrow deposition parameter window. We find that the pure $Bi_2FeCrO_6$ phase is formed within a narrow window around an oxygen pressure $PO_2 = 1.2 \times 10^{-2}$ mbar and around a substrate temperature $T_S$=680 °C. At lower temperature or higher pressure, $Bi_{7.38}Cr_{0.62}O_{12+x}$ _also called (b*$Bi_2O_3$) and $Bi_2Fe4O_9$ /$Bi_2(Fe,Cr)_4O_{9+x}$ phases are detected, while at lower pressure or higher temperature a $(Fe,Cr)_3O_4$ phase forms. Some of these secondary phases are not well known and have not been previously studied. We previously reported Fe/Cr cation ordering as the probable origin of the tenfold improvement in magnetization at saturation of our $Bi_2FeCrO_6$ film, compared to $BiFeO_3$. Here, we address the effect of the degree of cationic ordering on the magnetic properties of the $Bi_2FeCrO_6$ single phase. Polarization measurements at room temperature reveal that our Bi2FeCrO6 films have excellent ferroelectric properties with ferroelectric hysteresis loops exhibiting a remanent polarization as high as 55–60 $\mu C/cm^2$ along the pseudocubic (001) direction.

**I. INTRODUCTION**

Multiferroic materials, in which a magnetic and a ferroelectric order coexist, have attracted much attention because of their fundamental interest and their potential use in technological applications. Unfortunately, multiferroic materials are rare in nature and none of them have been used yet in practical applications due to either their too small response to an external field or their very small coupling between electric or magnetic properties. Materials that are multiferroic at *room temperature* (RT) are even scarcer. An example of such multiferroic material is $BiFeO_3$ (BFO).[1,2] Bulk $BiFeO_3$ intrinsically shows weak ferromagnetism and ferroelectricity at RT. Although this compound is an antiferromagnet, the magnetic moment of the canted antiferromagnetically ordered spins give rise to a weak ferromagnetism. The ferroelectricity, on the other hand, originates from a non-centrosymétriques crystal structure, in which the bismuth lone pair plays a crucial role in the large structural distortion.[3,4] During the past few years, special attention has been devoted to $BiFeO_3$ thin films mainly because of few reports of significantly enhanced ferroelectric and enhanced magnetic properties with respect to those of the bulk.[2,5] The intrinsic ferromagnetic properties of BiFeO3 were exhaustively debated and it is now accepted that the magnetization value of pure BFO is low.[6] The large number of sometimes contradictory studies of the multiferroic properties in BFO thin films has nevertheless boosted the research on multiferroic materials and their potential applications. Motivated by the report[7] of single atomic layer superlattice ordering in double perovskite $La_2FeCrO_6$, Baettig *et al.*[8] proposed an analogous Bi-based compound, namely $Bi_2FeCrO_6$ (BFCO), as a potential multiferroic material having a saturated magnetization superior to that of BFO. Their density functional calculations for the (111)-ordered structure indicated an $R3c$ symmetry for the ground state with a polarization of 80 mC/cm$^2$ along the (111) direction, and a ferrimagnetic ordering with a magnetization of 160 emu/cm$^3$ (corresponding to $2\mu_B$/ f.u.) and a magnetic transition at ~100 K. Ferromagnetic materials are mostly metallic and therefore not suitable to sustain ferroelectricity. Insulating magnetic materials, on the other hand, tend to show an antiferromagnetic coupling between their magnetic ions; hence, the interest of inspecting ferrimagnetic materials as possible candidate for RT multiferroic materials. Although there has been considerable progress on the growth of a variety of double perovskites,[9,10] work on Bi-based double layered perovskites is still in its infancy. They belong to the family of oxides with general formula $Bi_2BB'O_6$ with alternating $BiBO_3$–$BiB'O_3$ perovskite blocks. Since the major field of applications of this kind of materials is expected to be in thin film tunneling devices, there is a special interest in the preparation of very high quality thin films. The most promising materials are Bi-based double perovskites with *B*-site cations having different valency, such as $Bi_2MnNiO_6$, which has been synthesized so far only at high pressure.[10] We recently reported the first epitaxial synthesis of another Bi-based double perovskite, namely BFCO films obtained by pulsed laser deposition (PLD). The most striking characteristic of these BFCO films is their good multiferroic properties *at* RT.[11] The PLD-grown epitaxial BFCO films were deposited directly on (100)-oriented SrTiO3 (STO) single crystalline substrates[12] and on STO coated with an epitaxial SrRuO3 buffer layer. Our results confirmed the existence of multiferroic properties in BFCO films that were predicted by Baettig and Spaldin's *ab initio* calculations.[11] The existence of a magnetic ordering at RT in BFCO epitaxial PLD-grown thin films is an unpredicted yet very promising result that needs to be further investigated. The $Bi_2FeCrO_6$ films have a double perovskite crystal structure very similar to that of BFO and exhibit a clear Fe–Cr ordering along the (111) direction.[13]

In this paper, we analyze in detail the conditions for growing phase pure epitaxial films of BFCO by PLD on (100)-oriented STO. The influence of the deposition conditions on the PLD-grown thin films in a wide range of deposition temperature $T_S$ (600–850 °C) and ambient oxygen pressure $PO_2$ ($10^{-4}$ to $10^{-2}$ mbar) has been investigated. The phase stability of the BFCO films, a crucial factor for its integration into functional nanostructures, has also been studied.

## II. DEPOSITION AND CHARACTERIZATION EXPERIMENTAL SETUPS

BFCO films 90 nm thick were deposited by PLD on (100)-oriented SrTiO3 substrates. We used an excimer KrF ($\lambda$=248 nm) laser at a repetition rate of 8 Hz and with a fluence on the target of ~2 J cm$^{-2}$. A dense PLD ceramic target composed of a mixture of 50 mol % $BiFeO_3$ and 50 mol % $BiCrO_3$ was used. The distance from the target to substrate was 55 mm. Under these conditions, the film growth rate, estimated from thicknesses measured by x-ray reflectometry, was from ~1.7 to 2.0 Å s$^{-1}$. The films were prepared under several oxygen atmospheres, from vacuum (i.e., $10^{-6}$ mbar in our experimental setup) up to $1 \times 10^{-1}$ mbar and at substrate temperatures ranging from 600 to 850 °C. After deposition, the samples were cooled down to RT under the same oxygen pressure as the one used for deposition. The crystal structure of the films was studied by x-ray diffraction (XRD) using a Panalytical X'Pert Pro diffractometer.

Chemical composition and valencies of the magnetic cations were investigated by x-ray photoelectron spectroscopy (XPS). The analysis was performed at RT at a base pressure of ~$10^{-9}$ mbar with an ESCALAB 220i-XL system using monochromatic Al *K*$\alpha$ (1486.6 eV) radiation as excitation source (full width at half maximum of the Ag $3d_{5/2}$ line=1 eV at 20 eV pass energy). The reported binding energies were referred to C 1*s* line at 284.5 eV. The area of analysis was 4 mm$^2$. Complementary analysis using scanning electron microscopy (SEM) was used to characterize the film microstructure [secondary electrons (SEs)] and composition [backscattering electrons (BEs)]. Surface morphology was analyzed using a DI-Enviroscope AFM (Veeco) atomic force microscope (AFM). Magnetization hysteresis loops were measured at RT with the magnetic field oriented in-plane using a vibrating sample magnetometer (VSM) from ADE Technologies. A maximum magnetic field of 10 kOe was applied parallel to film plane, and then decreased down to −10 kOe by step of 500 Oe, and back up to 10 kOe. Around zero, the field steps were decreased down to 50 Oe for better resolution. An average factor of 20 measurements per applied magnetic field provided an absolute sensitivity of about 10−6 emu. The magnetic responses of the sample holding rods and of bare substrates _without film_ were also measured in order to confirm that the magnetic signal indeed originates from the thin films. The experimental error is mainly determined by the error in the estimation of the film volume. For some samples, the results were verified using a SQUID magnetometer from Quantum Design. RT polarization measurements were performed with a ferroelectric tester (aixACCT TF2000 analyzer).

## III. RESULTS AND DISCUSSIONS

### A. Optimization of the deposition parameters

#### *1. Deposition temperature*

Initially the films were deposited at fixed oxygen pressure ($PO_2$) of $1.2 \times 10^{-2}$ mbar and at deposition temperature (substrate temperature *TS*) ranging from 600 to 850 °C. Figure 1 compares the surface morphology and the phases developed in BFCO films for these various substrate temperatures.

We found the optimum growth temperature at 680 °C, for which the films present a smooth surface (rms roughness measured by AFM less than 2 nm) and show no secondary phases [cf. Fig. 1(a)]. A broad scan (20°–80°) for the film deposited at 680 °C [cf. Fig. 1(d), film B] shows only diffraction peaks from the substrate and (00*l*) pseudocubic reflections from the BFCO layer, which reveals that the films grow epitaxially with the *c* axis out of plane. A *c*-axis oriented tetragonal structure of the BFCO thin films is promoted by the STO substrate during the PLD deposition, owing to a smaller lattice constant of the substrate and a larger thermal expansion coefficient than that of the film. The cube-on- cube epitaxial relationship between BFCO film and STO substrate has been confirmed by phi-scan measurements around the (103) cubic reflection of STO (not shown).

In addition, reciprocal space map around (204) cubic reflection _not shown here_ exhibit only one spot for BFCO (204), which is aligned with that of the STO substrate. These results suggest that the BFCO film has a pseudo-tetragonal unit cell. In contrast, films deposited at lower temperatures [600 °C, Fig. 1(b)], have a rough surface covered by square outgrowths having ~250 nm in lateral size and ~30 nm in height. The corresponding XRD spectrum for these films [cf. Fig. 1(d), film A] exhibits additional reflections, which we attribute to the presence of Bi-rich phases. These phases segregate to the surface in self-organized arrays of squared shaped outgrowths. The films surface deposited at a temperature higher than the optimum, i.e., at 850 °C, presents elongated structures [Fig. 1(c)] and several additional peaks are observed in their XRD spectra [Fig. 1(d), film C]. We identified the two main secondary phases present to be $Bi_2(Fe,Cr)_4O_9$ (BFC*) and $Bi_2O_3$. The secondary phases labeled BFC* on Fig. 1(d) have never been reported (thus no patterns found in the XRD pattern database). We assumed, however, that their structure is similar to the $Bi_2Fe_4O_9$ (BF*) parasitic phase of BFO, as well as to similar layered bismuth-transition metal oxide systems, such as $Bi_2Fe_{4-x}Ga_xO_9$ and $Bi_2Fe_{4-x}Al_xO_9$.[14] These kinds of compounds present an orthorhombic crystal structure close to that of $Bi_2Fe_4O_9$ (Ref. 15) and show an antiferromagnetic order with Néel temperature depending on the $x$ ratio but generally well below RT.

Analyzing the XRD spectra around the pseudocubic 002 reflection of BFCO, we also notice a change in the BFCO out-of-plane lattice parameter with the growth temperature. As seen in Fig. 1(e), the peak shifts toward lower angles with increasing deposition temperature, suggesting that the $c$-axis of the BFCO unit cell increases with the deposition temperature. Additionally, reciprocal space maps around (103) reflection of STO (not shown here) revealed that for all films the BFCO phase is fully strained and have an in-plane lattice parameter very close to that of STO (~0.390 nm).

### 2. Deposition pressure

XRD patterns and corresponding microstructures of the films grown at a fixed substrate temperature (namely 680 °C) and at different oxygen pressure $PO_2$ are presented in Fig. 2. (00$l$) peaks (symbol red open diamonds) in the XRD spectra [Fig. 2(a)] corresponding to the 001-oriented $Bi_2FeCrO_6$ phase are clearly visible near the STO substrate reflections (labeled S). We found an optimum oxygen pressure of $PO_2=1.2\times10^{-2}$ mbar, for which the films grown appear to be single phase. These BFCO peaks are not detected in the film deposited under vacuum, suggesting that the BFCO phase is either present in very small quantity or not formed. In contrast, the peaks detected match very well some intense reflections of metallic Fe or iron oxide phases (FO), such as the reflections of a-$Fe_2O_3$ (hematite) or g-$Fe_2O_3$ (maghemite), or the reflections of spinel $(Fe,Cr)_3O_4$ or mixed sesquioxide $(Fe,Cr)_2O_3$ phases (FCO). The absence of any pure chromium oxides peak reflections in the XRD pattern has to be noted, suggesting that the films probably crystallize preferentially into the FCO phases.

For films deposited at an oxygen pressure between $1\times10^{-4}$ and $6.0\times10^{-3}$ mbar, we observed a mixture of $Bi_2FeCrO_6$ and FCO phases.

The change in the dominant phase in the films reflects the change in the amount of Bi incorporated in the films due to the high volatility of metallic Bi. Indeed, decreasing the oxygen pressure during the deposition, the Bi content in the film decreases, while at high pressure of $O_2$, the XRD spectra contain only major reflections corresponding to b, d, or b*Bi2O3 (BO) phases or a mixture of them. At oxygen pressure higher than the optimal value (e.g., $2.7\times10^{-3}$ mbar) reflection peaks corresponding to the FCO phases are not visible anymore but additional peaks appear in the (27°–33°) 2q range. We identified these reflections as originating from b*$Bi_2O_3$ and b or d-$Bi_2O_3$ phases for the films grown at $PO_2=2.7\times10^{-2}$ and $1\times10^{-1}$ mbar, respectively. The phases grown at the highest oxygen pressure exhibit a rough surface with typical BO square-shaped outgrowths protruding out of the surface.

SEM micrographs of the films deposited at various pressures are shown in Fig. 2(c)–2(e). The film deposited under vacuum [Fig. 2(c)] exhibits rectangular grains. The films grown at low oxygen pressure ($4.2\times10^{-3}$ mbar) exhibit the same rectangular features on the surface but with a lower density and a different composition and structure, possibly a
spinel with composition $(Fe,Cr)_3O_4$ (see Sec. III below). The Bi deficiency of the spinel parasitic phase outgrowths obtained at low pressure was confirmed by SEM analysis of the same area in the BE mode. Upon deconvolution of the XRD peaks of film 2, we compute a fraction of 37% of this secondary phase [Fig. 2(b)]. We note here that the intensity of the peak corresponding to FCO increases as the oxygen pressure decreases.

### 3. Chemical analysis

XPS has been used to investigate the chemical composition and the oxidation states of the Fe and Cr elements present in the different BFCO layers. For the films deposited at the optimal pressure of $1.2\times10^{-2}$ mbar and at various temperatures, the XPS analysis of the surface of the sample labeled "A" on Fig. 1 (i.e., a film deposited at 600 °C) reveals that the square-shaped outgrowths are Bi-rich, Cr deficient and Fe-poor. Their overall chemical composition corresponds to that of $Bi_{7.38}Cr_{0.62}O_{12+x}$ (Cr-doped $Bi_2O_3$ or b*-Bi2O3) or $Bi_{14}CrO_{24}$, which have a tetragonal crystal structure with $a=b=7.7596$ Å and $c=5.7388$ Å,[16] and $a=b=8.672$ Å and $c=17.21$ Å,[17] respectively.

The XPS analysis also shows that the oxidation state is 3+ for each cation. The fact that the outgrowths are Bi-rich was further confirmed by the SEM image in the backscattered electron mode (not shown). Since the oxidation state of Cr in $Bi_{14}CrO_{24}$ has been reported to be 4+,[17] and that $Cr^{4+}$ ions could not be evidenced on the XPS spectrum of the Cr 2p XPS line of our films, we can exclude the presence of this phase in the layer. Chemical analysis of the film deposited under vacuum confirmed the strong deficiency in bismuth that was inferred from XRD analysis and gives a Fe/Cr cationic ratio close to unity.

For chromium, the Cr 2p XPS spectra revealed that the oxidation state is $Cr^{3+}$ for all the films. For iron, as it is well known, the Fe 2p core level splits into 2p1/2 and 2p3/2 components. The binding energy of Fe 2p3/2 is expected to be 710.7 eV for Fe3+ and 709 eV for Fe2+. To quantify the fraction of iron in each chemical state using XPS, we used the method described by Aronniemi et al.[18] Using the traditional Shirley background subtraction between 700 and 740 eV, we deconvoluted the Fe 2p core level for the different samples. The line shape used to represent the 2p mains peaks was a Gaussian–Lorentzian (GL) product with a constant exponential tail. We imposed the fitting parameters such that the tail parameters and the GL ratio of the 2p1/2 main peak are equal to those of 2p3/2 and the satellites are purely Gaussian and without any tail.

For the films deposited at 680 °C (the optimal temperature) and at various pressures, a detailed analysis of the peaks of samples 1–3 (Fig. 3) was done, fitting two iron-oxygen clusters with different Fe2+ and Fe3+ valencies. We find that the binding energy of the main peak at around 711.0 eV (in film 3) is shifted to a lower value (within the instrument precision of 0.1 eV) of ~710 eV in sample 1. In XPS spectrum of sample 1, the intensity of 2p3/2 satellite (around 719 eV) characteristic of the $Fe^{3+}$ ions in $Bi_2FeCrO_6$ becomes weak and less resolved due to the rising intensity of a peak at about 716 eV. This latter peak is normally assigned to the satellite for the $Fe^{2+}$ ions, analogous to the spectrum of wustite, FeO.[19,20] The contribution of the Fe2+ in the films is gradually reduced with increasing oxygen pressure. The Fe state in sample 1 can indeed be modeled as the sum of $Fe^{2+}$ and $Fe^{3+}$, which gives rise to the $2p3/2$ and $2p1/2$ main peaks and to four satellites. The Fe $2p3/2$ main peak maximum of the $Fe^{2+}$ component has a binding energy of 709 eV, while that of the Fe3+ lies around 710.6eV. The atomic concentrations giving the best fit of the XPS spectra are 21% Fe (with 52% $Fe^{2+}$ and 48% $Fe^{3+}$), 20% Cr ($Cr^{3+}$), 1% $Bi^{3+}$, and 57% $O^{2-}$. This composition corresponds to that of the spinel $Fe^{3+}[Fe^{2+}Cr^{3+}]O4$. The oxidation states of the cations in this spinel have been investigated previously by Yearian *et al.*[21] in order to study the influence of Fe/Cr ratio. In our measurements, however, we cannot assess the site occupancy of each ion. This would require a complementary investigation, such as x-ray absorption spectroscopy or x-ray circular magnetic dichroism. We can notice here that the very existence of $Fe^{2+}$ rules out the presence of $\alpha$-$Fe_2O_3$ and $\alpha$-$(Fe,Cr)_2O_3$ phases in the film (which contain only $Fe^{3+}$ cations). We thus conclude that the phase formed in the films 1 and 2 is $(Fe,Cr)_3O_4$ (FCO). In the film 2, the Fe component is a mixture of 17% $Fe^{2+}$ and 83% of $Fe^{3+}$, whereas for film 3, $Fe^{2+}$ cannot be detected within the measurement error.

Due to the lack of oxygen during deposition, the valence fluctuation of Fe is actually expected. The most important spectral parameter values obtained as a result of the fitting are presented in Table I. Comparing the data with results obtained on BFO thin films shown in the last column, we conclude that the film 3 (or the film B in Fig. 1) prepared at $PO_2$ of $1.2 \times 10^{-2}$ mbar is nearly stoichiometric $Bi_2FeCrO_6$ while the film 1 deposited under vacuum is close to stoichiometric $(Fe,Cr)_3O_4$. We also compare the Fe $2p$ XPS spectra obtained from BFO (Ref. [12]) and BFCO 90 nm thick thin films. The chemical state of iron in 90 nm thick BFO is quantified using XPS and the result is listed in Table I for comparison. We found that the oxidation state for iron is 3+ for both BFCO and BFO and that the ions occupy the octahedral sites only (excluding the presence of maghemite, exhibiting both Fe in octahedral and tetrahedral sites).

However, we find that the XPS Fe $2p$ spectrum in the BFCO film has two different features compared with that of BFO film. First, the relative intensity of the satellite peak at 719 eV compared to the main $2p3/2$ peak ($I_{sat}/I_{main}$) is reduced, and second, the $2p3/2$ main peak is slightly shifted to a lower binding energy. As it is known, the p-d hybridization state between oxygen and transition metal orbitals considerably affect the satellite intensities.[22,23] Quantitatively, the intensity of the satellite peak is approximately given by the following equation $1 - T^2/\Delta(Q-\Delta)$, where $T$ is the ligand $2p$ to metal $3d$ hybridization energy, $\Delta$ is the ligand $2p$ to the metal $3d$ charge transfer energy, and $Q$ is the Coulomb interaction between the core hole and the $3d$ electrons.[23] Therefore the satellite intensity is strongly influenced by the ligand $2p$ to metal $3d$ hybridization energy $T$. Because $\Delta$ and $Q$ cannot reasonably reproduce the simultaneous changes in both the satellite intensities and their positions, we believe that the decreased satellite intensity in BFCO compared to BFO is mainly caused by an increase in *T*. Reporting Harrison's relations,[24] *T* between Fe 3*d* and O 2*p* varies with the interatomic distance as $r^{-3.5}$. Additionally, the different lattice mismatch between the films and substrate induces anisotropic strains at the interfaces,[25] leading to a change in the interatomic distance in the thin films. Nevertheless, the increase in *p-d* hybridization energy *T* of BFCO film implies that the Fe–O bonds in BFCO are more covalent than in BFO, causing the intra-atomic Coulomb energies *Q* and *U* to be screened more effectively by the density of the polarization of the bonds.[23]

## *4. Phase diagram*

The analysis of the various phase formation under different growth conditions presented in the previous section is graphically summarized in the phase diagram presented in Fig. 4. As usual with the deposition of Bi-based compounds,[26,27] the high volatility of Bi brings complexity to the pressure-temperature phase diagram and makes the optimization process a difficult task. The BFCO single phase is obtained only in a narrow window, around $1.2 \times 10^{-2}$ mbar and 680 °C. Several elements point toward a relationship between nonstoichiometry defects (i.e., oxygen and cation vacancies) and the destabilization of the BFCO single phase.

The presence of *M*-rich phases (*M*=Fe or Cr) in the films deposited at low pressure of oxygen and for substrate temperatures higher than the evaporation temperature of bismuth [$T_S > T_{evap}(Bi)$], together with a high concentration of cationic *A*-site vacancies in the $A_2BB'O_6$ perovskite structure, are the consequence of the evaporation of a large fraction of non-oxidized bismuth or of the decomposition of $Bi_2O_3$ unstable at high temperature.[27] On the contrary, at low temperature and high $O_2$ pressure, the oxidation of bismuth is favored, forming the Bi2O3 stable phase. In this case, the adatoms have a low mobility on the surface of the substrate, which would produce interstitials oxygen atoms in the film. Actually, the perovskite framework allows oxygen vacancies but not oxygen interstitials. The oxygen surplus in the perovskite structure leads to the formation of cation vacancies.[28] Therefore, this nominal increase of the oxygen content compared to the correct stoichiometry, results in a large amount of cation vacancy in the perovskite phase. The cation vacancies in *A*-sites and insufficient mobility at the surface explain both the reduction of the BFCO cell volume and the formation of stable bismuth oxide. Finally, at high temperature and high pressure, the BFCO single phase is not stable and decomposes into different phases. The smaller unit cell volume of BFCO thin films (with respect to that of the bulk[8]) and the decreasing of the intensity of the XRD BFCO(004) peak at very high temperatures suggest that the oxygen vacancies concentration is not the principal factor causing the XRD BFCO(004) peak shift to lower angle with temperature increase.

As shown in Fig. 1(e), the (002) peak intensity of BFCO deposited at 850 °C decreases compared to that of the film deposited at 680 °C. This is due to the nucleation of secondary phases at the expense of the BFCO single phase.

The oxygen or cation non-stoichiometry-induced defects thus probably favors the decomposition of the BFCO phase into more stable phases such as $Bi_2(Fe,Cr)_4O_9$ and $b^*Bi_2O_3$. At the growth temperature of 680 °C, the BFCO lattice is compressed in-plane at the interface due to the misfit in lattice constant between film and substrate ($a_p$=0.3932 and 0.3905 nm, respectively).

For the 90 nm thick film, this compressive strain stabilizes the single phase of BFCO. We further looked into the possible effect of the Bi evaporation on the pure BFCO phase formation. Figure 4(b) shows the central region of the temperature-pressure phase diagram, under which conditions promote the growth of pure BFCO together with the Bi vapor pressure curve obtained from the literature[29] on the same plot. Not surprisingly, there is a strong correlation between them, suggesting a strong influence of the evaporation of Bi on the stabilization of BFCO single phase within the studied temperature range.

## B. Magnetic properties of BFCO films

### *1. Origin of the enhanced magnetic properties in BFCO single phase films*

Several reasons could explain the change in the macroscopic magnetization in BFCO films. The increased magnetization may be attributed to a suppressed inhomogeneous antiferromagnetic spin structure and to an increased canting angle. These contributions are usually occurring in a stressed structure and lead only to the occurrence of weak magnetic moment in the otherwise AFM phase,[30] which cannot be the main factors in our case.

One other reason for the strong magnetization of 90 nm thick BFCO films (sample B of Fig. 1 or sample 3 of Fig. 2) is that local ferromagnetic supercells might exist as a consequence of double perovskite structure with Fe–Cr cationic order. It is known that a strong ferromagnetic coupling exists between $Fe^{3+}$ and $Cr^{3+}$ under 180° superexchange interactions.[31] As we previously reported,[11] there is a net improvement in macroscopic magnetization in BFCO thin films at RT compared with BFO thin films so BFCO and BFO would exhibit two different types of magnetic exchange interaction. Magnetic hysteresis loops with low saturation magnetization were observed in $Bi(Fe_{1-x}Cr_x)O_3$ bulk compounds at RT,[32] and it was found that the addition of Cr in BFO enhanced the magnetization, a fact explained by the formation of a locally ordered structure.

In pure BFO, an Fe($d5$-$d5$) antiferromagnetic coupling between spins gives rise to a weak ferromagnetism via a slight canting of the otherwise antiferromagnetic spin structures.[33] For BFCO, instead of a Fe–O–Fe bond, we have to consider the Fe–O–Cr interaction. From the theoretical prediction, the magnetic moment expected for BFCO depends of the type on Fe–O–Cr interactions. For ferromagnetic interactions (where the Fe and Cr spins are parallel) involving high spin (HS) $Fe^{3+}$($d5$)-O–$Cr^{3+}$($d3$), a total magnetization corresponding to $4\mu_B$ per site (or $8\mu_B$ par Fe–Cr Pair) is predicted. For antiferromagnetic interactions (antiparallel spins for the HS $Fe^{3+}$($d5$) and Cr3+($d3$) ions), a magnetization value of $1\mu_B$ per site ($2\mu_B$ per Fe–Cr pair) would be expected for the HS state of Fe. Our BFCO films exemplifies a $d5$($Fe^{3+}$)-$d3$($Cr^{3+}$) system because (i) both Fe and Cr are formally in the +3 ionic states and (ii) the existence of a cationic ordering in the films has been revealed by XRD reciprocal space mapping (RSM).[13,34] (in contrast, Kim et al.[35] obtained an epitaxial disordered solid solution of BFO and BCO).

Magnetic properties of BFCO thin films deposited in the optimal conditions (sample B of Fig. 1 or sample 3 of Fig. 2) are shown in Fig. 5(a). The magnetization versus magnetic field $M(H)$ hysteresis curves were measured with a SQUID magnetometer at 300 and at 10 K with an applied magnetic field of 10 kOe and show no significant depen- dence on temperature. The saturation magnetization is ~145 electromagnetic units per cubic centimeters (emu/cc), which corresponds to about 0.96 Bohr magneton ($\mu_B$) per cation site or ~$1.91\mu_B$ per Fe–Cr pair (i.e., ~$1.91\mu B$ per BFCO formula unit). From the predictions of the Structure Prediction Diagnostic software (SPUDS),[36] we found an average angle for the Fe–O–Cr bond of about 175°. Thus, the slight difference between the theoretical and experimental value of magnetization is probably due to the deviation of this bond angle from 180° (which was used for the ab initio predictions). In our case, the macroscopic moment of BFCO seems to be well explained by a ferrimagnetic order.

To further investigate the improved magnetism in BFCO, we plot in Fig. 5(b) the temperature dependence of the magnetization, measured in a magnetic field of 500 Oe applied parallel to the surface of the substrate. No magnetic transition temperature was detected up to 400 K, the temperature limit of the magnetometer. At 400 K, the film is still magnetic, meaning that the Curie temperature of the BFCO film investigated is higher than 400 K. At low temperature, the variation of spontaneous magnetization ($M_S$) with temperature ($T$) is given by the Bloch law[37] deduced from the spin wave model

$$M_S(T) = M_S(0)(1 - AT^{3/2}),$$

where $M_S(0)$ is the spontaneous magnetization at 0 K and $A$ is a constant that depends on the exchange integral $J$ between the neighboring spins ($A \sim 1/J^{3/2}$). In the low-temperature limit, the dominant factor known to suppress magnetization is spin-wave excitations. It is clear from the data in Fig. 5(b) that the magnetic transition temperature exceeds 300 K, as suggested by Kamba et al.[38] The positive value of the exchange integral $J$ confirms the presence of ferrimagnetic ordering in the BFCO film investigated.

The much higher $T_c$ observed in our BFCO films, compared with the $T_c$ predicted by *ab initio* calculations may be due to the large strain that changes the Fe–O–Cr interatomic distances and/or bonding angles, consequently reducing the exchange energy among spins, and shifting the Curie temperature to higher values. In BFCO epitaxial films, the small octahedra tilting and the variations of the lattice parameters relative to the predicted bulk will induce a decrease of the Fe–O and Cr–O bond lengths from 2.05 and 1.98 Å in the bulk to 1.97 and 1.95 Å, respectively, in the films under compressive strain. The Fe–O bond in the BFCO film is 2% smaller that those in the BFO film. The change in Fe–O interatomic distance is confirmed by the change in the characteristic binding energy extracted from the position of the XPS Fe $2p$ line for BFCO with respect to the binding energy obtained for BFO. However, this change in bond lengths is insufficient to explain alone a fivefold increase of the magnetic transition temperature.

The ferromagnetic superexchange interaction is also governed by the polarization effects. In fact, the presence of $6s^2$ lone pair electrons at the $Bi^{3+}$ ion sites in BFCO can lead to the formation of long-range molecular electric field, as in $BiMnO_3$, $Bi_2NiMnO_6$, and Bi2CoMnO6.[39–41] The built-in molecular electric field originating from the large anisotropic overlap of these Bi $6s2$ lone pair electrons with the surrounding hybridized Fe $3d$-O $2p$ orbitals can facilitate their coupling to the long-range spin order through the minimization of the magnetoelastic energy.
This effect, therefore, would not only allow the direct coupling between the magnetization and the polarization field but it could also drastically enhance the ferromagnetic superexchange interactions between magnetic cations, leading to a magnetic Curie temperature close to the ferroelectric transition. Recent infrared and magnetic characterization of 200 nm thick BFCO film seems to confirm this fact.[38] RT ferroelectric properties of 200 nm thick films deposited under optimum conditions {red star in the phase diagram, Fig. 4(a)] are shown in Fig. 5(c). BFCO films show ferroelectric property with a remnant polarization of 55–60 $\mu C/cm^2$ along the (001) direction (after subtracting the leakage current contribution).

### *2. Further improvement of the magnetic properties induced by the cationic ordering*

Although the optimal deposition window to grow phase pure BFCO thin films is small, good quality BFCO films with good magnetic properties could be grown at various pressures provided the suitable temperature is chosen accordingly. Several phase pure BFCO thin films have been prepared covering the range of deposition parameters allowed, as illustrated in the diagram phase of the Fig. 4. The corresponding RT *M*(*H*) magnetic hysteresis loops obtained by VSM, with *H* in the plane of the substrate, are shown in Fig. 6(a).

We found that even though all the films were phase pure BFCO, their magnetic hysteresis and saturation magnetization varied significantly. Better magnetic properties were obtained for the films prepared at high *T* and high oxygen pressure (i.e., P5 and P6) compared to those prepared at low *T* and low oxygen pressure (i.e., P1 and P2). XRD patterns clearly indicate that all BFCO films were epitaxially grown and no evidence of presence of secondary phases was observed for any of them. For all the films, the RSM study around the (111) STO cubic reflection revealed the presence of the superlattice reflections (111) and (333) (in the pseudocubic "STO-like" single perovskite unit cell indexing), indicating a clear *B*-site ordering in the doubled perovskite unit cell _with rhombohedral lattice parameter (7.94 Å). The measurement of the intensity of (111) superstructure revealed a higher intensity, hence a clear reduction of the density of antisite defects for high temperature and high-pressure grown films. It is not yet clear whether the improved *B*-site cation ordering is uniquely related to a better structural quality induced by a better kinetics at higher temperature but from the magnetic results it is clear that the films grown at higher temperature and higher pressure show a higher spontaneous magnetization value $M_S$ compared with lower temperature and lower pressure ones. In Fig. 6(b), $M_S$ is plotted as a function of degree of Fe/Cr ordering estimated from the superlattice reflection to the normal reflection intensity ratio $I_{111} / I_{222}$. It is clearly seen that the increase of $M_S$ is very strongly correlated with the increase in the degree of *B*-site cation ordering. This relationship between the magnetization and the degree of *B*-site ordering observed in BFCO films can be explained by a ferro/ferrimagnetic coupling within Fe–O–Cr structure.

## IV. CONCLUSION

In summary, we have grown epitaxial BFCO films onto STO(100) substrates, and explored the impact of the deposition pressure and substrate temperature on the phases which form upon crystallization. We found that the growth of pure BFCO thin films is strongly favored in a relatively narrow window around to 680 °C and $1.2\times10^{-2}$ mbar of oxygen.

Within this window, we have found that epitaxial BFCO films grown at $1.2\times10^{-2}$ mbar are single-phase for thickness up to at least 250 nm. For lower oxygen pressure, all the films contain $(Fe,Cr)_3O_4$ [or $(FeCr)_2O_3$], secondary phases.

Higher oxygen pressure induces the formation of $Bi_2O_3$, along with destabilization and decomposition of the $Bi_2FeCrO_6$ phase to form tetragonal $Bi_2O_3$ and $Bi_2(Fe,Cr)_4O_9$ phases. The presence of these secondary phases and the absence of cationic order within the remaining BFCO phase lead to a significantly lower magnetization in the films. From the XPS spectra of Fe $2p$ in single phase BFCO, we established that only Fe in the oxidation state 3+ is present without any evidence of Fe2+ within a resolution of few at. %. Compared to BFO films, we found a significant difference in the satellite structure of the Fe $2p$ XPS spectra, suggesting a different hybridization state of Fe $3d$–O $2p$. The XPS Fe $2p$ line of BFCO had a smaller satellite intensity compared to that of BFO, which indicates a larger Fe $3d$ to O $2p$ hybridization in BFCO.

Magnetic measurements revealed a maximum for the magnetic moment of $\sim 1\mu_B$ per site (corresponding to $2m_B$ per Fe–Cr pair, or per formula unit) at 10 K obtained for 90 nm thick film, almost identical with that obtained at 300 K. This value is explained in terms of *ferrimagnetic* superexchange magnetic interaction which exists in BFCO unlike the *canted spin structure of antiferromagnetic* BFO. The BFCO system offers an example of $d$5-$d$3 magnetic superexchange interaction because both Fe and Cr are ordered and formally in the +3 ionic states. Further investigations are under way, especially to determine the effect of film thickness, epitaxial strain and Fe–O–Cr bond angles on the magnetic properties of BFCO thin films.

**ACKNOWLEDGMENTS**

One of the authors (R.N.) wishes to thank Professor Karen Kavanagh and Sarmita Majumder from Simon Fraser University, British Columbia, for high resolution XRD measurements. Part of this research was supported by NSERC and CFI (Canada) and FQRNT (Québec). One of the authors (P.F.) acknowledges the support of CIfAR.

**TABLE I**. Spectral parameters of the fit of the Fe 2p3/2 core-level XPS spectra for BFCO films 1 and 3 shown in Fig. 3. Tabulated are the binding energy of the 2p3/2 main peaks maximum (BE), the half width, the intensity ratio of the satellite to the 2p3/2 main peak, and the separation between them. The corresponding spectral parameters for the fit of the Fe 2p3/2 peak of a BFO film are only presented for comparison.

| Sample | FCO (1) | | BFCO (3) | BFO |
|---|---|---|---|---|
| Iron valency | $Fe^{2+}$ | $Fe^{3+}$ | $Fe^{3+}$ | $Fe^{3+}$ |
| Be $2p_{3/2}$ (eV) | 709.0 | 710.6 | 710.9 | 711.3 |
| $2p_{3/2}$ half width $\beta$ (eV) | 1.45 | 1.8 | 1.8 | 1.7 |
| Be $2p_{1/2}$-Be $2p_{3/2}$ (eV) | 13.2 | 13.6 | 13.4 | 13.4 |
| $2p_{1/2}$ half width $\beta$ (eV) | 1.5 | 2.0 | 2.0 | 2.1 |
| $2p_{1/2}/2p_{3/2}$ intensity ratio | 0.52 | 0.55 | 0.54 | 0.54 |
| $2p_{3/2}$ satellite shift (eV) | 6.7 | 8.8 | 8.6 | 8.2 |
| $2p_{3/2}$ sat/$2p_{3/2}$ intensity ratio | 0.36 | 0.17 | 0.21 | 0.25 |

**Figure Captions**

**Figure 1.** (Color online) SEM micrographs [(a)–(c)] and corresponding XRD patterns [(d) and (e)] of BFCO films grown under $1.2 \times 10^{-2}$ mbar of oxygen at different substrate temperatures: (a) 680 °C (film B), (b) 600 °C (film A), and (c) 800 °C (film C). The bright squares on the surface correspond to b*Bi2O3 (b*BO) outgrowths. (d) Broad θ-2θ scans of the same films reveal pure BFCO phase only in films grown at 680 °C while (e) a zoom around the 004 reflection shows the dependence of the growth temperature on the peak position. The symbols (circles and diamonds) correspond to Kβ and WLα peaks from the STO substrate. BFC*, labels correspond to $Bi_2(Fe,Cr)_4O_9$ phases.

**Figure 2.** (a) XRD patterns for four films grown at 680 °C and at different pressures (from bottom to top, vacuum (1): $PO_2=4.2 \times 10^{-3}$ mbar (2); $PO_2=1.2 \times 10^{-2}$ mbar (3) and $PO_2=2.7 \times 10^{-2}$ mbar (4), $PO_2=1.0 \times 10^{-1}$ mbar (5). Note here that the sample 3 and the sample B of Fig. 1 are prepared in the same conditions. The peak around 2_ =40° corresponds to STO substrate contamination. BFC _open diamonds_, b_BO, BO, and BFC_ labels correspond to BFCO, b*$Bi_2O_3$, α/β $Bi_2O_3$ and $Bi_2(Fe,Cr)_4O_9$ phases, respectively. Solid diamonds, stars, and plus symbols correspond to Kb, WLa, and STO substrate contaminations. (b) High resolution XRD obtained for a small range of 2_ angles for the film deposited under a pressure of oxygen of $4.2 \times 10^{-2}$ mbar (e). [(c)–(e)] SEM images showing several microstructures for the films 1, 3, and 5 respectively.

**Figure. 3.** Observed XPS Fe 2p core-level spectra of the samples deposited under vacuum (1), at $PO_2=4.2 \times 10^{-3}$ mbar (2) and at $PO2=1.2 \times 10^{-3}$ mbar (3), respectively, showing the presence of both $Fe^{2+}$

and $Fe^{3+}$ valencies for the films 1 and 2 and of the single valency $Fe^{3+}$ for the film 3 deposited under the optimal deposition conditions. The parameters used for the fit are listed in Table I.

**Figure 4.** (a) Pressure-temperature phase diagram for thin films of the Bi–Fe–Cr–O system with a nominal thickness of 90 nm. (b) (oxygen pressure-substrate temperature) data points in the phase diagram for which pure phase BFCO films are obtained together with the vapor pressure curve of metallic Bi drawn in the same diagram, evidencing the critical importance of Bi volatility in obtaining phase pure BFCO films.

**Figure 5**. (a) Magnetization hysteresis curves for a BFCO thin film on STO (100) in the field range of 500 Oe at 10 and at 300 K measured by SQUID magnetometer. (b) Temperature dependence of magnetization of BFCO thin film on STO (100) measured in a 0.5 kOe (0.05 T) magnetic field applied parallel to the substrate surface. The magnetic field is applied parallel to the film [i.e., In the (100) direction of STO]. (c) Current vs applied voltage curve and the corresponding ferroelectric hysteresis loop and obtained at RT for a 200 nm thick BFCO thin films. The solid line corresponds to the polarization hysteresis loop after subtracting the leakage contribution.

**Figure 6.** (a) Magnetic hysteresis loops of various BFCO thin films prepared with different deposition conditions but all prepared with deposition conditions within the phase diagram region where single phase BFCO is obtained, as indicated by the open circles labeled P1 to P6 in the phase diagram. In the inset, a zoom around the origin. (b) Relationship between the degree of ordering in Fe/Cr estimated from the $I_{111}/I_{222}$ ratio and the saturated magnetization measured at RT (Applied magnetic field is in-plane). The inset shows RSM-extracted profile lines around the (111) superstructure reflection obtained for the different films.

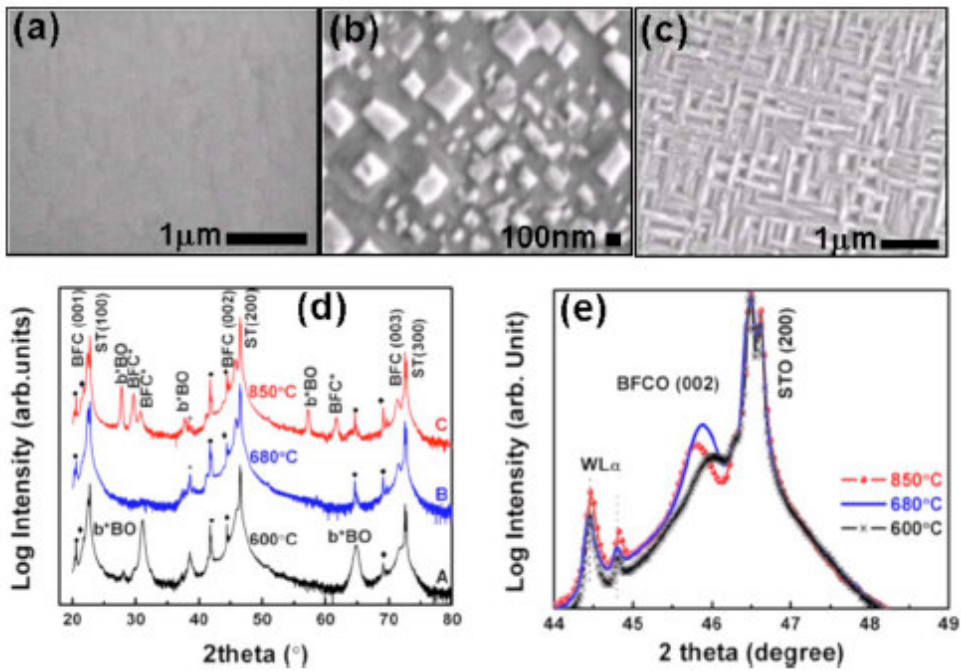

Figure 1 R. Nechache et al.

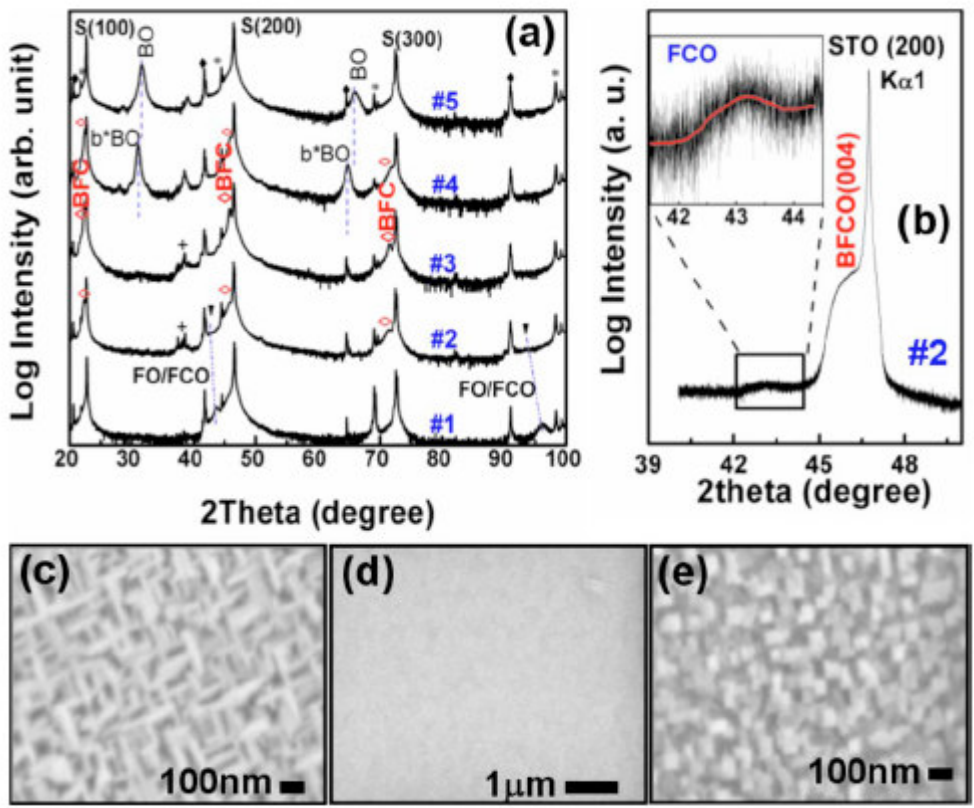

Figure 2 R. Nechache et al.

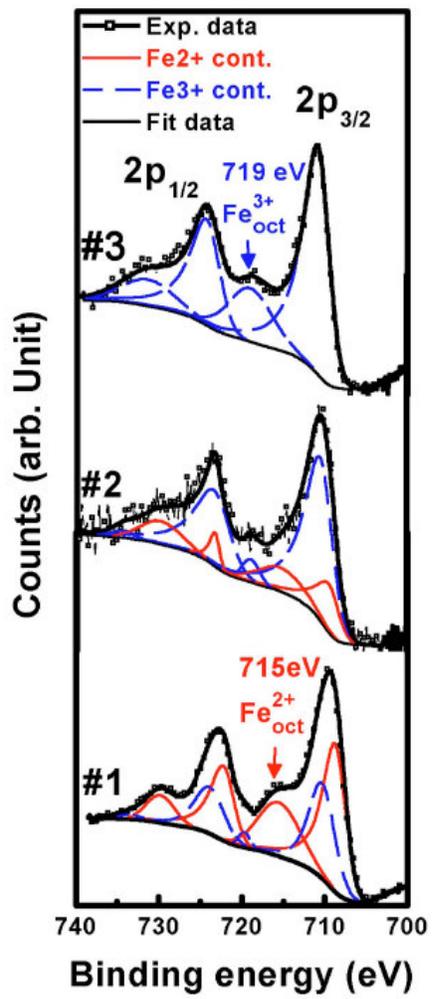

Figure 3 R. Nechache et al.

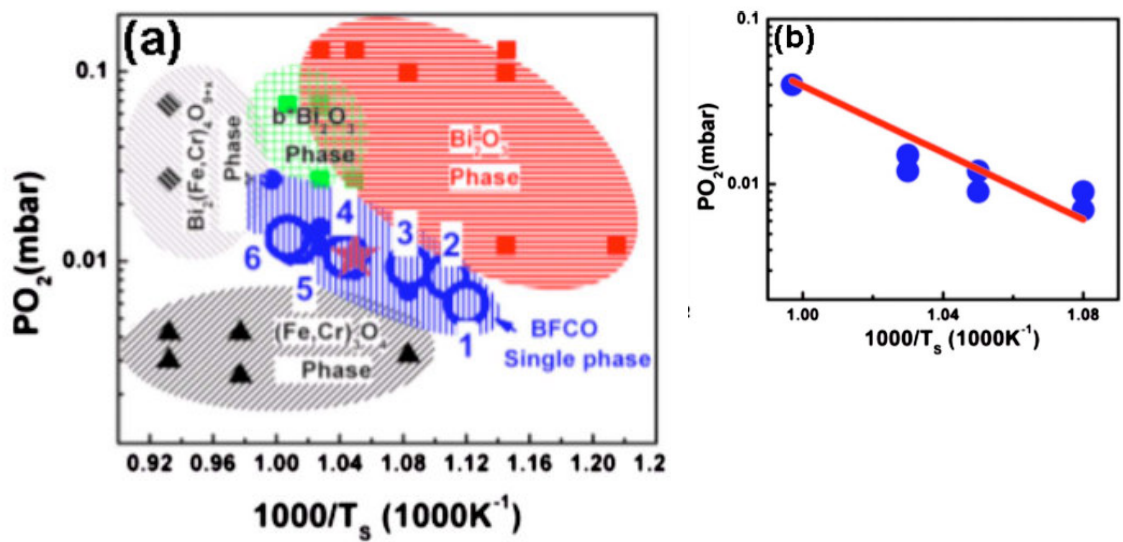

Figure 4. R. Nechache et al.

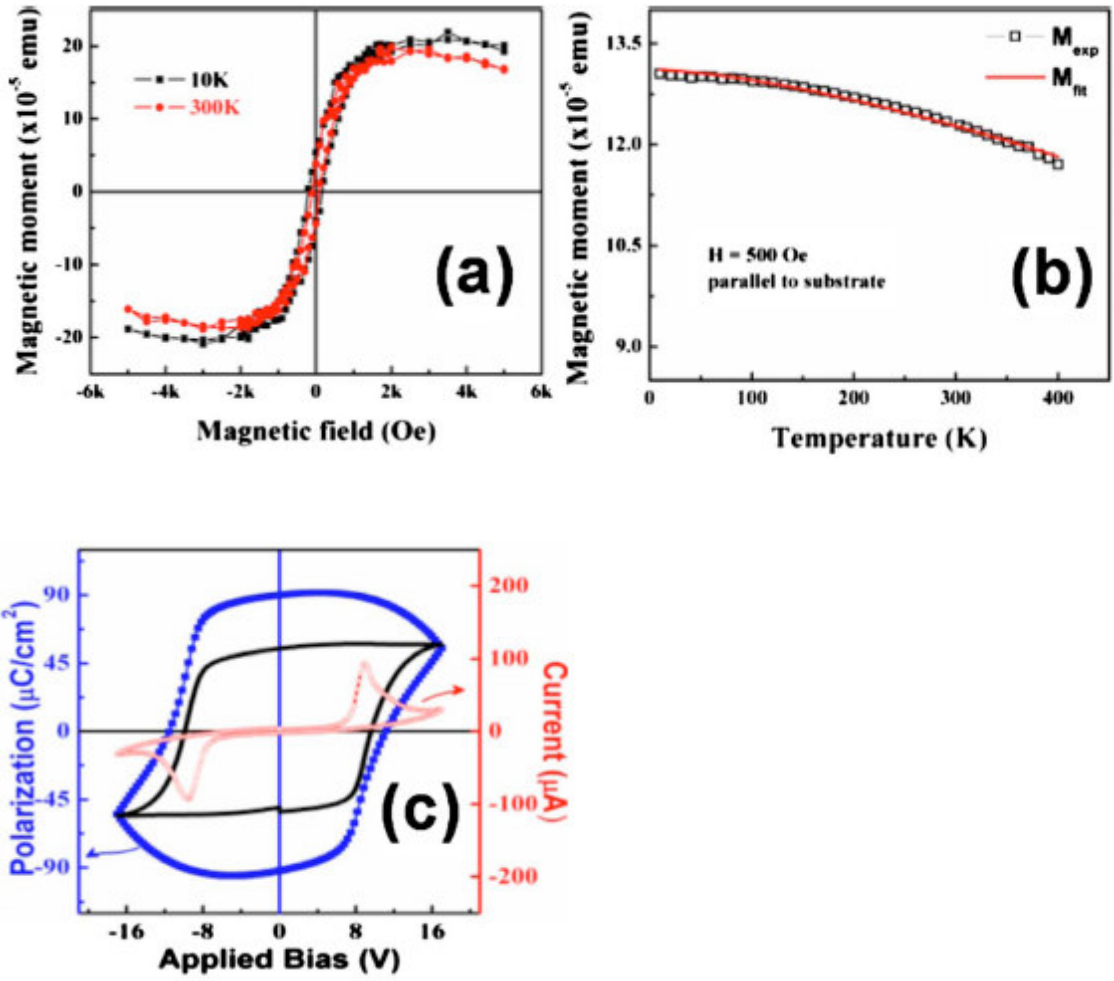

Figure 5. R. Nechache et al.

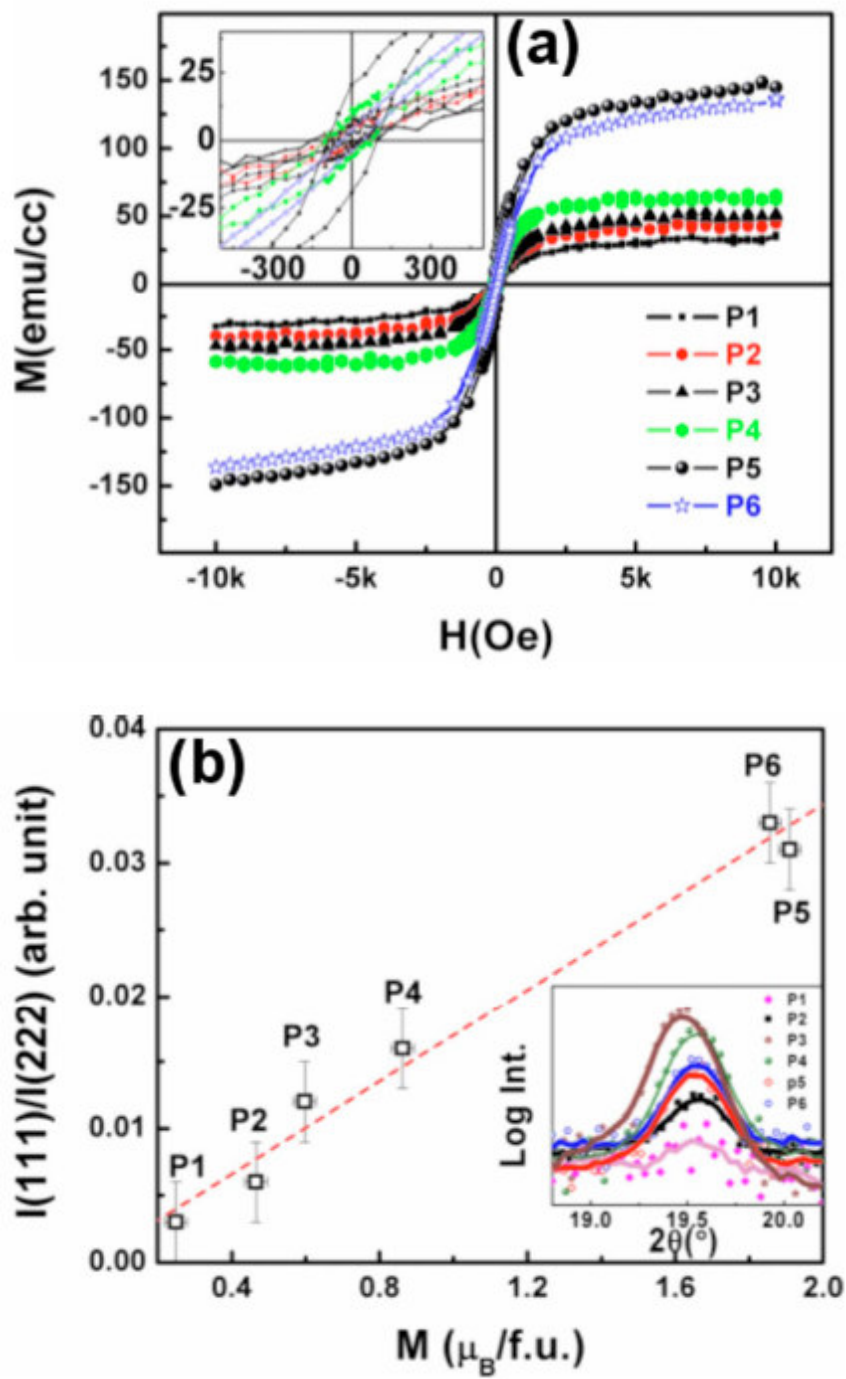

Figure 6. R. Nechache et al.